# Decomposable multiphase entropic descriptor

D. Frączek[1, *], R. Piasecki[2, *]

[1] *Department of Materials Physics, Opole University of Technology, Katowicka 48, 45-061 Opole, Poland*
[2] *Institute of Physics, University of Opole, Oleska 48, 45-052 Opole, Poland*

A B S T R A C T

To quantify degree of spatial inhomogeneity for multiphase materials we adapt the entropic descriptor (ED) of a pillar model developed to greyscale images. To uncover the contribution of each phase we introduce the suitable "phase splitting" of the adapted descriptor. As a result, each of the phase descriptors (PDs) describes the spatial inhomogeneity attributed to each phase-component. Obviously, their sum equals to the value of the overall spatial inhomogeneity. We apply this approach to three-phase synthetic patterns. The black and grey components are aggregated or clustered while the white phase is the background one. The examples show how the valuable microstuctural information related separately to each of the phases can be obtained at any integer length scale. Even dissimilar hidden statistical periodicities can be easily detected for chosen phases built-up of compact regular clusters.

## 1. Introduction

Amongst many factors, the microstructure of multiphase materials particularly influences the prediction of their effective physical properties [1, 2]. One of the simplest configuration features of a random microstructure is the degree of spatial inhomogeneity at different length scales. A quantitative characterization needs detailed information about phase occupation of very small sub-domains of a given media. This kind of information is quickly accessible by mean of digitized micrographs, which should be representative for considered material, for instance, its surface or cross-sections layers. Recently introduced entropic descriptor (ED) for greyscale images employs models with different types of pillars: decomposable and non-decomposable [3]. Despite simplicity of the basic idea, the entropic descriptors can be applied even for a statistical reconstruction of complex materials [4-5]. In this letter we present an adapted ED based on a model of non-decomposable pillars. Their different heights play a role of distinct shades of grey attributed to given phases. As we show later on, the adapted ED possesses a nice property: it can be splitted into individual *phase descriptors* (PDs). Each of them allows for quantification of the spatial inhomogeneity separately for respective phases. Therefore, the multiscale analysis of spatial inhomogeneity making use of PDs is undoubtedly of some importance within the field of searching connections between structure–property relations of random multiphase materials [6-7].

---

[*] *E-mail addresses*: d.fraczek@po.opole.pl (D. Frączek), piaser@uni.opole.pl (R. Piasecki)
 *Daniel Frączek is a recipient of a Ph.D. scholarship under a project funded by the European Social Fund.*

## 2. The multiphase entropic descriptor and its splitting into phase descriptors

Let us consider a system's pattern of size $L \times L$ (in pixels) of pillars, where each of them is treated as the whole entity (non-decomposable) of a fixed height '$i$'. In general, a given pattern can be sampled by $\lambda(k) = [(L-k)/z + 1]^2$ cells of size $k \times k$ with a sliding factor $1 \leq z \leq k$ provided $(L-k) \bmod z = 0$. Here, the $z = 1$ is chosen that gives the maximal overlapping of the cells. In fact, we analyse auxiliary $L'(k) \times L'(k)$ patterns, called further 'maps', where $L'(k) \equiv (L - k + 1) k$. Those maps composed of the sampled cells placed in a non-overlapping manner can be treated as the representative ones since they clearly reflect at every length scale $k$ the morphological features of the initial digitised images. Then, the actual macrostate AM($k$) for a representative map can be described with the help of a set $\{m_{i\alpha}(k)\}$, $i = 1, 2, …, w$, and $\alpha = 1, 2, …, \lambda(k)$, where $0 \leq m_{i\alpha}(k) \leq k^2$ are the multiplicities of pillar appearance of height '$i$' inside $\alpha$th sampling cell.

For the case we consider, every cell is fully occupied. Hence, the general formula given by Eq. (A1) in [3] describing the number $\Omega(k)$ of realizations (microstates) for any AM($k$) macrostate considerably simplifies. It can be rewritten in a slightly modified notation as

$$\Omega(k) = \prod_{\alpha=1,…,\lambda(k)} \left( \frac{k^2!}{\prod_{i=1,…,w} m_{i\alpha}(k)!} \right). \tag{2.1}$$

It is a simple observation that the pillars of fixed integer height can play the role of $1 \times 1$ (in pixels) parts of corresponding phase component in a multiphase material build of $w$ phases. Thus the integers $m_{i\alpha}(k)$ can be treated as $\alpha$th cell occupation numbers attributed to the $i$th phase at length scale $k$. Applying further this interpretation, any $m_{i\alpha}(k)$ indicates the number of finite size objects (pixels) of $i$th phase component inside $\alpha$th cell of size $k \times k$. Here, for a digitised image the pixels play the role of smallest portions or grains of a given phase.

For auxiliary pattern at every length scale $k$, the basic constraints for the occupation numbers of $i$th phase, $i = 1, 2, \ldots, w$, are described by

$$\sum_{\alpha=1,\ldots,\lambda(k)} m_{i\alpha}(k) = M_i(k) \tag{2.2}$$

where $M_i(k)$ stands for the total number of pixels of $i$th phase. Hence, the auxiliary $k$-dependent corresponding concentration $\varphi_i(k)$ can be written as

$$\varphi_i(k) = M_i(k) \big/ \lambda(k) k^2 \tag{2.3}$$

Since we consider $w$-phase material, at every length scale $k$ we have $0 < \varphi_i(k) < 1$ with lack of porosity condition $\sum_i \varphi_i(k) = 1$. Further, to simplify notation we will omit the length scale $k$ wherever it does not leads to misunderstanding. The above approach allows us to compute the configurational entropy $S(k) = \ln \Omega(k)$, with Boltzmann constant taken as unity for convenience,

$$S(k) = A(k) - \sum_{i=1,\ldots,w} \sum_{\alpha=1,\ldots,\lambda(k)} \ln[m_{i\alpha}(k)!] \tag{2.4}$$

Here, the term $A(k) \equiv \lambda(k) \ln(k^2!)$ has been obtained after performing double summation.

The construction of a multi-phase entropic descriptor needs evaluation of statistical dissimilarity (kind of a 'distance') of the actual map from the reference maximally uniform one. Therefore, the 'experimental' and 'theoretical' maps should be compared for every length scale $k$. The maximum theoretical value $S_{\max}(k) = \ln \Omega_{\max}(k)$ of the system's entropy is accessible for the multiphase reference macrostate $RM_{\max}(k)$ that corresponds to the highest possible degree of uniformity. Then, one can find that each of the $r_i$ cells is occupied by $m_{i,0} + 1$ of pixels of corresponding $i$th phase while the rest $\lambda - r_i$ cells by $m_{i,0}$ only. Thus, the reference macrostate $RM_{\max}(k)$ can be described with the help of a set $\{(\lambda - r_i)m_{i,0}(k) + r_i(m_{i,0}+1)\}$. In this case the simple relations hold: the number $r_i = (M_i \bmod \lambda) \in \{0, 1, ..., \lambda - 1\}$ and the average cell occupation $m_{i,0} = (M_i - r_i)/\lambda \in \{0, 1, ..., k^2 - 1\}$. This kind of multiphase configuration appears at given length scale $k$, when for every $i$th phase and any pair of different cells $\alpha \neq \beta$ the following inequalities are fulfilled $|m_{i\alpha}(k) - m_{i\beta}(k)| \leq 1$. If we have $r_i = 0$ and each $m_i = m_{i,0} \equiv M_i/\lambda$ then a completely uniform distribution of '$w$' phases appears. In Appendix, to get clear insight into the recipe for filling up the cells, we present at fixed length scale the appropriate macrostates for a three-phase surrogate pattern.

For $i$th phase, the $w$-tuples attributed to the $RM_{max}$ contain those values of occupation numbers, which belong to the corresponding set $\{m_{i,0}, m_{i,0}+1\}$. Thus, after rearranging and grouping the occupation numbers in $\ln \Omega_{max}(k)$ one can obtain for $S_{max}(k)$ the compact formula

$$S_{max}(k) = A(k) - \sum_{i=1,...,w}(\lambda - r_i)\ln(m_{i,0}!) - \sum_{i=1,...,w} r_i \ln(m_{i,0}+1)! \qquad (2.5)$$

Again, a sequence of the summations is interchanged and the impact of different phases is put on the first place.

A reasonable quantitative comparison, at different length scales $1 \leq k \leq L$, of the statistical dissimilarity of AM and $RM_{max}$ macrostates needs averaging of the difference $S(k)_{max} - S(k)$ over the number $\lambda(k)$ of sampling cells. At this stage, the multiphase entropic descriptor quantifying the degree of the overall inhomogeneity per cell is written in a form used previously for different models [8-11]

$$S_\Delta(k) = (S_{max} - S)/\lambda, \qquad (2.6)$$

extended also in [12] to the formula employing of Tsallis entropy [13].

The "$S_\Delta$ splitting problem" raised by one of us (DF) consists in asking if this multiphase entropic descriptor can be decomposed into phase descriptors in such a way that their sum over $w$-phases equals exactly to the overall $S_\Delta$. One of possible approaches requires rewriting of the term $A(k) \equiv \lambda(k)\ln(k^2!)$ as two equivalent forms, one for the $S(k)$ and second for the $S_{max}(k)$. With the condition $\sum_i \varphi_i(k) = 1$, the first of them is

$$A(k) = \sum_{i=1,\ldots,w} \lambda\, \varphi_i(k) \ln(k^2!) \tag{2.7a}$$

Now, Eq. (2.4) can be rewritten as a sum of $i$th phase non-negative contributions (the proof of the inequality $f_i(k) \geq 0$ is elementary)

$$S(k) = \sum_{i=1,\ldots,w} [\lambda \varphi_i \ln(k^2!) - \sum_{\alpha=1,\ldots,\lambda(k)} \ln(m_{i\alpha}!)] \equiv \sum_{i=1,\ldots,w} f_i(k) \tag{2.7b}$$

The second equivalent form for the term $A(k)$

$$A(k) = \sum_{i=1,\ldots,w} [(\lambda - r_i)\varphi_i \ln(k^2!) + r_i \varphi_i \ln(k^2!)] \tag{2.8a}$$

applied to Eq. (2.5) gives consequently

$$S_{\max}(k) = \sum_{i=1,\ldots,w} \left\{ \begin{array}{l} (\lambda - r_i)[\varphi_i \ln(k^2!) - \ln(m_{i,0}!)] + \\ r_i [\varphi_i \ln(k^2!) - \ln(m_{i,0}+1)!] \end{array} \right\} \\ \equiv \sum_{i=1,\ldots,w} f_{i,\max}(k) \tag{2.8b}$$

where $f_{i,\max}(k) \geq f_i(k) \geq 0$. According to Eq. (2.6), the final form of phase component-separated ED reads

$$S_\Delta(k) = \sum_{i=1,\ldots,w} (f_{i,\max} - f_i)/\lambda \equiv \sum_{i=1,\ldots,w} f_{i,\Delta}(k). \tag{2.9}$$

One point needs a brief explanation. The above notation suggests the phase descriptors ($f_{i,\Delta} \equiv$ PDs) can formally be named as the phase *entropic* descriptors if we accept an extended definition of $\Omega_i$ and $\Omega_{i,\max}$ for $i$th phase, that is

$$\Omega_i(k) = \frac{(k^2!)^{\lambda \varphi_i}}{\prod_{\alpha=1,\ldots\lambda(k)} m_{i\alpha}!} \tag{2.10}$$

and

$$\Omega_i(k) = \left(\frac{(k^2!)^{\varphi_i}}{m_{i,0}!}\right)^{\lambda - r_i} \left(\frac{(k^2!)^{\varphi_i}}{(m_{i,0}+1)!}\right)^{r_i} \tag{2.11}$$

However, it looks somewhat untypical because now both $\Omega_i$ and $\Omega_{i,\max}$ are usually non-integer quantities. Therefore, we prefer use the notion phase descriptors.

One remark is in order. Another entropic descriptor of statistical complexity introduced in [14] can be also easily defined in a similar way to multiphase media. However, it cannot be split into phase components and only the overall statistical complexity relates to multiphase materials. Note the present approach can be also adapted for specific series consisted of limited number of different "components" like, for instance, DNA sequence. In addition, its extension to three-dimensional case can be easily done.

## 3. Illustrative examples

We check our approach by applying it to a few aggregated/clustered systems with black, grey, and white components.

**Example 1.** First, we examine the phase components, $f_{1,\Delta}$, $f_{2,\Delta}$, and $f_{3,\Delta}$, of adapted entropic descriptor for a simple three-phase pattern of size $126 \times 126$ in pixels, see the inset in Fig. 1. The pattern is composed of nine $42 \times 42$ domains which are the replicas oriented differently of the subdomain extracted from [15] and used previously in [4]. Despite of equal phase concentrations one can observe differences in their aggregations. The irregular black phase clusters are the largest ones. Thus, the question what phase influences mostly the overall spatial inhomogeneity, thick solid black line in Fig. 1, is a simple one in this example. Indeed, the first component $f_{1,\Delta}$, thin solid black line, related to the black phase gives the biggest contribution at all length scales. The component $f_{2,\Delta}$ related to the grey phase, thin solid grey line, and respectively, the $f_{3,\Delta}$ coming from the white phase, thin solid line blue online, provide the minor contributions. This descriptive example clearly shows that phase descriptors enable for quantitative evaluation at different length scales some characteristic features of displacement for each of the phases. For instance, the maxima of $f_{1,\Delta}$ and $f_{2,\Delta}$ appearing at same scale $k_{\max}(\text{black, grey}) = 10$ and correspond roughly to the bigger average size of those irregular clusters while for the white phase the related maximum is shifted on the left and $k_{\max}(\text{white}) = 8$.

**Example 2.** In this example we investigate *simultaneous* aggregation of the two phases, black and grey. It depends on neighbourhood rules, used by a simple cellular automata, that is von Neumann or Moore type consisting of all four nearest neighbours (n.n.) or eight (n.n. + n.n.n.) ones, respectively. The third white phase is treated as a background one. The starting three-phase pattern is a random one, see the inset in Fig. 2a. For illustration purposes, we show only two stages of an evolution of the aggregation performed for two combinations of the neighbourhood rules: (i) Moore type for black and von Neumann for grey pixels called as *mixed rule* (MvN); (ii) only von Neumann type for black and grey pixels named as *doubled rule* (vNvN).

The rules apply when two pixels, neighbouring by a side or a corner (the one of the eight directions is picked randomly), are a pair of white and black pixels, or white and grey one. This is a first necessary condition to make the interchange of positions of white with black pixel or white with grey. Using a white-black pair to present more details, the second condition consists in checking if the following inequality holds: $nn_0 \leq 2$, where $nn_0$ is a number of black n.n. around the black centre pixel. In this way, we do not destroy of already existing black clusters larger than three pixels. If yes, the third condition of majority comes to play that relates to the numbers of black neighbours, {$N$(white centre; MvN) and $N$(black centre; vNvN)} in case (i) or {$N$(white centre; vNvN) and $N$(black centre; vNvN)} in case (ii). For instance, the following inequality condition, $N$(white centre; vNvN) $\geq$ $N$(black centre; vNvN), causes the positions interchange of white and black centres. Similar procedures apply for a white-grey pair.

Fig. 2a-d shows all PDs as a function of length scale $k$. The solid (dashed) lines of a phase corresponding colour relate to doubled (mixed) rules used. In addition, the two stages of evolution of aggregation measured by total number of accepted interchanges for grey and black phases are illustrated. The results of different combinations of the neighbourhood rules are shown in Fig. 2a for the stage 1 (about $2.65 \cdot 10^6$ acceptations) and in Fig. 2b for the stage 2 (about $7.94 \cdot 10^6$ acceptations). In turn, Fig. 2c shows the results for the two different stages of aggregation performed with the use of mixed rule only and similarly, Fig. 2d for only doubled rule applied. To discern more details all insets are exemplary $120 \times 120$ subdomains of the main patterns of size $360 \times 360$.

One can observe from Fig. 2a-d that the mixed rule is much less effective than doubled one in a creation of bigger aggregations of black and grey phases at lower scales. In addition, the Moore rule promotes a creation of black clusters in comparison to von Neumann rule used for grey phase. However, at larger scales this trend changes and usually the spatial distribution of grey phase is slightly more inhomogeneous.

**Example 3.** We consider now the problem of detecting of hidden statistical periodicities for chosen phases in exemplary $360 \times 360$ patterns generated randomly by our computer program. Fig. 3a-b shows three phase configuration 'A' of 128 regular black clusters and configuration 'B' of 576 grey clusters composed of 137 and 37 pixels, respectively. The white component is always treated as a background phase. For the two patterns, essential differences in arrangements are hardly seen at first view. However, the first of them, i.e. 'A', contains dissimilar hidden periodicities since only in this case the two additional and different rules were used to positioning the black and grey clusters, see Fig 3c. Specifically, each cell of $45 \times 45$ in size contains exactly two black clusters, the first rule, while four grey clusters occupy each of $30 \times 30$ cells, the second rule.

However, as Fig. 3d shows, the black and grey phase descriptors clearly reveal these differences. For example, an average distance equals to 44.6 for the sequence {50, 40, 43, 45, 48, 43, 43, 48, 42} of intervals between successive local minima of solid black curve. This suggests an appearance of a period close to the integer scale 45 in pixels. It corresponds to an approximate statistical periodicity in arrangement of black clusters. Even more precisely, for solid grey curve the corresponding average distance 29.9 indicates for an occurrence of statistical periodicity around the integer scale 30. In this case, the sequence of successive intervals is given by {30, 30, 29, 29, 31, 29, 30, 32, 31, 28}. Note such nearly regular intervals do not appear for a random configuration 'B'. One more remark is in order. The common minima for black and grey phases appear at scales around $k = 90$, 180 and 270, as expected.

## 4. Conclusions

In this paper we adapted overall entropic descriptor connected with greyscale images to the case of multiphase materials. Now, it quantifies the multiscale degree of average spatial inhomogeneity. To reveal contribution of any phase we propose the "phase splitting" of the adapted descriptor. In this way, formulas for each of the so-called phase descriptors describe separately the spatial inhomogeneity attributed to any phase-component of a system. The proposed phase descriptors appear to be a quite effective tool in materials science. The illustrative examples show that valuable microstructure features like phase dependent aggregations or hidden periodicities are detectable. This could make possible a searching of an impact of spatial arrangement of respective phases on effective properties of multiphase systems. This will be a subject of our future work.

**Appendix**

At length scale k = 3, we present an exemplary construction of auxiliary square pattern (map) for the toy pattern of size $L \times L = 4 \times 4$ composed of three phases: black, grey and white. We attribute to each $i$th phase the integers $m_{i\,\alpha}(k)$ being the occupation numbers of $\alpha$th cell by the phase-pixels. In this case we have $\lambda(k=3) = [L-k+1]^2 = [4-3+1]^2 = 4$ cells. Therefore, the corresponding actual macrostate, AM($k=3$), is defined as a set of occupation numbers $\{(m_{1\alpha}, m_{2\alpha}, m_{3\alpha})\}$, where $\alpha = 1, 2, 3$ and $4$ for each of the three phases, respectively. The detailed form of this macrostate reads $\{(4,4,1)_{\alpha=1}, (4,3,2)_{\alpha=2}, (3,4,2)_{\alpha=3}, (5,3,1)_{\alpha=4}\}$. This is in accordance with the map of side size $L' = k(L-k+1) = 3(4-3+1) = 6$ depicted below. The sampling procedure employs $3 \times 3$ cells, which are marked by thick frames (yellow online).

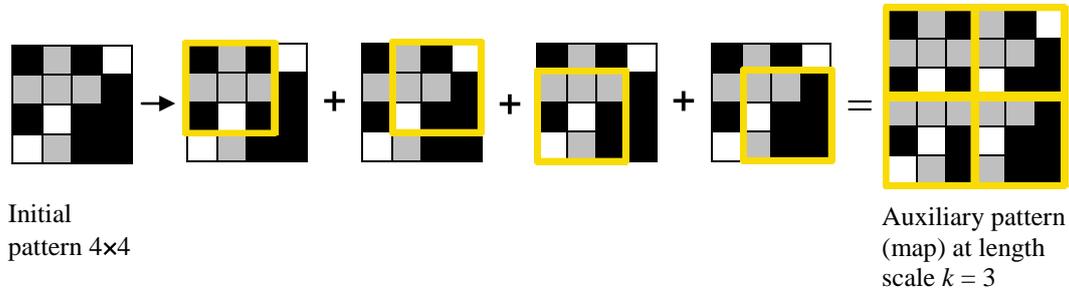

Initial pattern 4×4

Auxiliary pattern (map) at length scale $k = 3$

Applying Eq. (2.1) to the above map, the total number $\Omega(k=3)$ of realizations (microstates) for the actual macrostate, AM($k=3$), provides the simple formula

$$\Omega(3) = \left(\frac{3^2!}{4!4!1!}\right)\left(\frac{3^2!}{4!3!2!}\right)\left(\frac{3^2!}{3!4!2!}\right)\left(\frac{3^2!}{5!3!1!}\right). \tag{A1}$$

On the other hand, the reference macrostate, RM($k=3$), describes a multi-phase distribution of pixels corresponding to the maximal spatial uniformity. Such situation appears when for each $i$th phase and any pair of different cells, i.e. $\alpha \neq \beta$, the mono-phase occupation numbers satisfy the simple rule, $|m_{i\,\alpha}(k) - m_{i\,\beta}(k)| \leq 1$. The corresponding set can be written as $\{(4,4,1)_{\alpha=1}, (4,4,1)_{\alpha=2}, (4,3,2)_{\alpha=3}, (4,3,2)_{\alpha=4}\}$. Therefore, the total number $\Omega_{\max}(k=3)$ of realizations for the RM($k=3$) is given by the formula

$$\Omega_{\max}(3) = \left(\frac{3^2!}{4!4!1!}\right)\left(\frac{3^2!}{4!4!1!}\right)\left(\frac{3^2!}{4!3!2!}\right)\left(\frac{3^2!}{4!3!2!}\right). \tag{A2}$$

10In this case, the value of overall multiphase entropic descriptor, $S_\Lambda(k=3)$, equals to

$$S_\Lambda(3) = \frac{1}{\lambda}\ln\left(\frac{\Omega_{max}(3)}{\Omega(3)}\right) = \frac{1}{4}\ln\left(\frac{5!3!1!}{4!4!1!}\right) = \frac{1}{4}\ln\left(\frac{5}{4}\right) = 0.0558 \qquad (A3)$$

Finally, having employed Eq. (2.9), the $S_\Lambda(k=3)$ splits into the following phase descriptors:

$$f_{1,\Lambda}(3) = \frac{1}{4}\ln\left(\frac{5}{4}\right) = 0.0558, \qquad f_{2,\Lambda}(3) = 0, \qquad f_{3,\Lambda}(3) = 0. \qquad (A4)$$

The last two outcomes can be easily guessed since the corresponding maps have the possible maximal uniformity at scale $k=3$. At this scale, only the black phase gives a non-zero contribution, $f_{1,\Lambda} \geq 0$, to the overall spatial inhomogeneity. Therefore, its value must be same as for $S_\Lambda$.

**Figure captions**

**Fig. 1.** The multiphase entropic descriptor $S_\Delta$, thick solid black line, and its phase components $f_{1,\Delta}$ for black, $f_{2,\Delta}$ for grey, and $f_{3,\Delta}$ for white phase plotted as a function of length scale $k$ for a simple three-phase pattern of size $126 \times 126$ in pixels given in the inset. As expected, the first component $f_{1,\Delta}$, thin solid black line, related to the black phase gives prevalent contribution the overall spatial inhomogeneity at all length scales.

**Fig. 2.** The phase descriptors at two evolving stages as a function of length scale $k$ for simultaneous aggregation of two phases, black and grey. The descriptors make use of neighbourhood rules of 'mixed' type ≡ Moore for black and von Neumann for grey phase, dashed lines with phase corresponding colours, and 'doubled' kind ≡ von Neumann for both phases, solid lines; for details see the text. To perceive details all insets are exemplary $120 \times 120$ subdomains of the main patterns of size $360 \times 360$. (a) Stage 1 for mixed and doubled rules. (b) Stage 2 for mixed and doubled rules. (c) Stages 1 and 2 for mixed rule. (c) Stages 1 and 2 for doubled rule.

**Fig. 3.** Detection of dissimilar hidden statistical periodicities on example of two patterns, each of size $360 \times 360$, with subtle differences in their arrangement. (a) Surrogate configuration 'A'; (b) Random case 'B'. (c) Only in the case 'A' each cell of size $45 \times 45$ contains exactly two black clusters while four gray clusters occupy each of $30 \times 30$ cell. (d) The contribution of phase descriptors $f_{1,\Delta}$ and $f_{2,\Delta}$ related to black and grey phase, solid lines for the surrogate pattern 'A' while dashed lines correspond to random case 'B'. As expected, only in the case with hidden periodicities, i.e. 'A', appear quite regular intervals between the minima.

**Fig. 1**

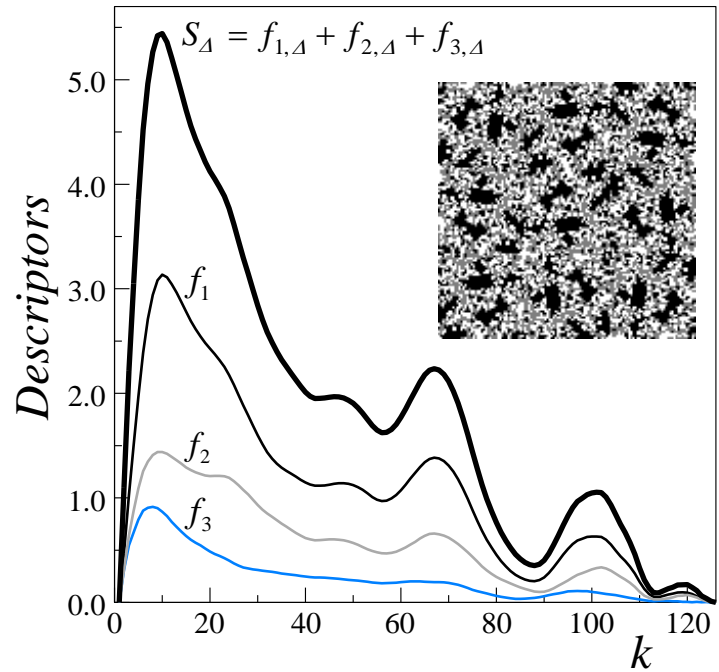

**Fig. 1.** The multiphase entropic descriptor $S_\Delta$, thick solid black line, and its phase components $f_{1,\Delta}$ for black, $f_{2,\Delta}$ for grey, and $f_{3,\Delta}$ for white phase plotted as a function of length scale $k$ for a simple three-phase pattern of size $126 \times 126$ in pixels given in the inset. As expected, the first component $f_{1,\Delta}$, thin solid black line, related to the black phase gives prevalent contribution the overall spatial inhomogeneity at all length scales.

**Fig. 2a**

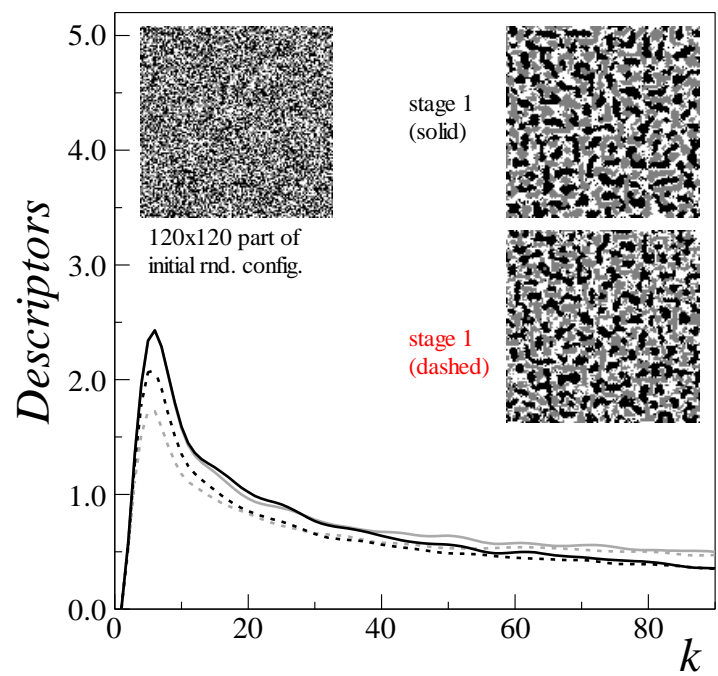

**Fig. 2b**

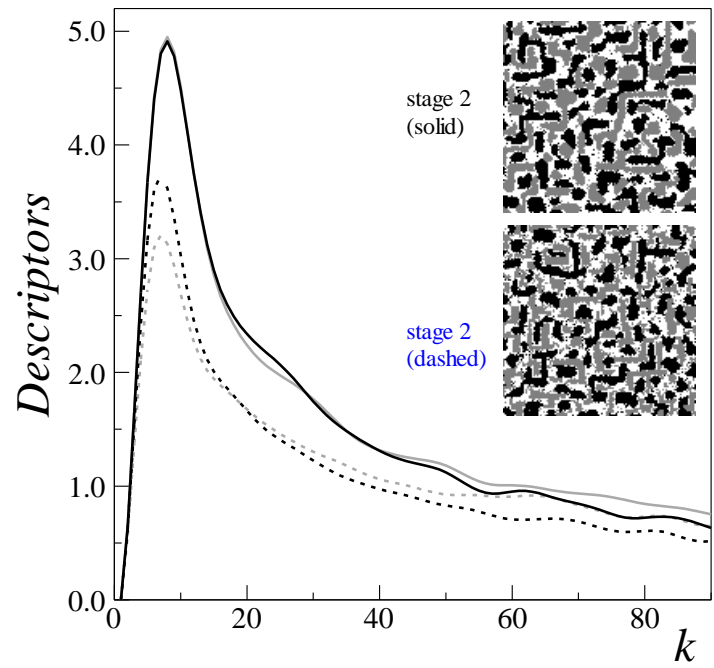

**Fig. 2c**

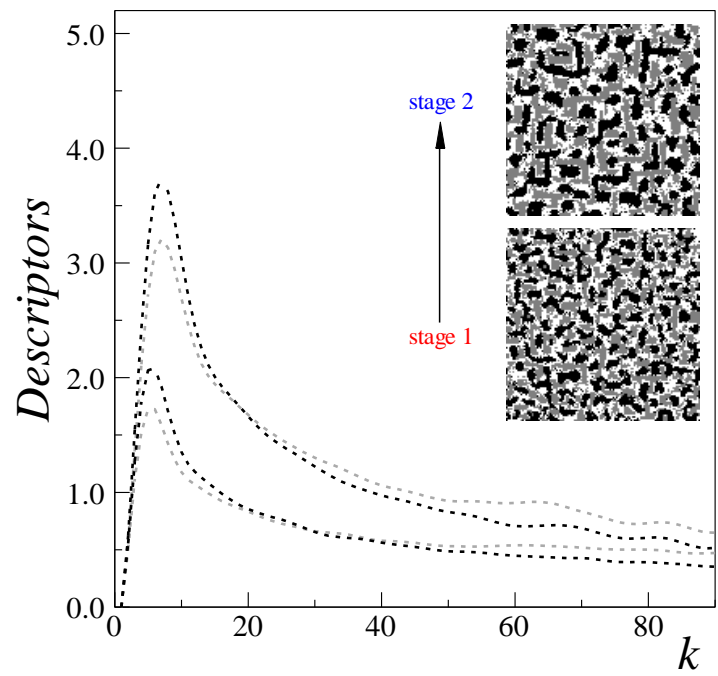

**Fig. 2d**

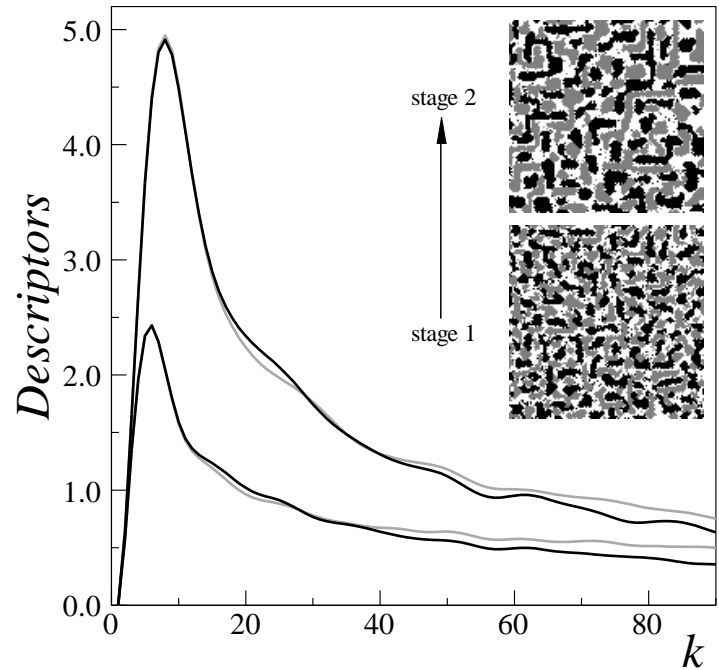

**Fig. 2.** The phase descriptors at two evolving stages as a function of length scale *k* for simultaneous aggregation of two phases, black and grey. The descriptors make use of neighbourhood rules of 'mixed' type ≡ Moore for black and von Neumann for grey phase, dashed lines with phase corresponding colours, and 'doubled' kind ≡ von Neumann for both phases, solid lines; for details see the text. To perceive details all insets are exemplary $120 \times 120$ subdomains of the main patterns of size $360 \times 360$. (a) Stage 1 for mixed and doubled rules. (b) Stage 2 for mixed and doubled rules. (c) Stages 1 and 2 for mixed rule. (c) Stages 1 and 2 for doubled rule.

**Fig. 3a**

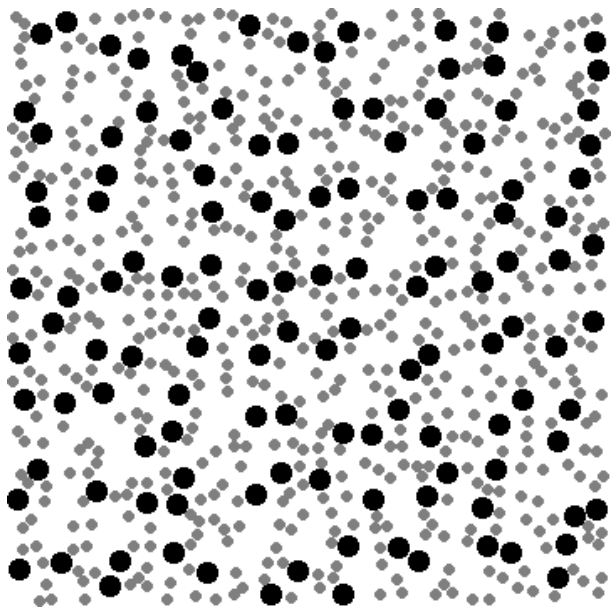

Configuration 'A'

**Fig. 3b**

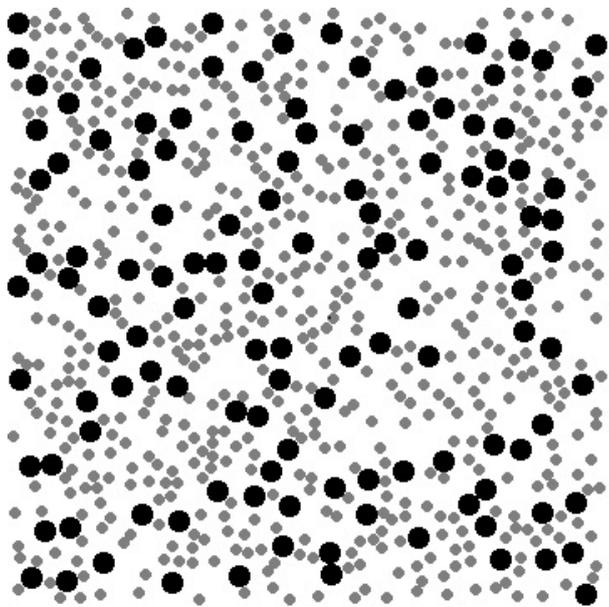

Configuration 'B'

**Fig. 3c**

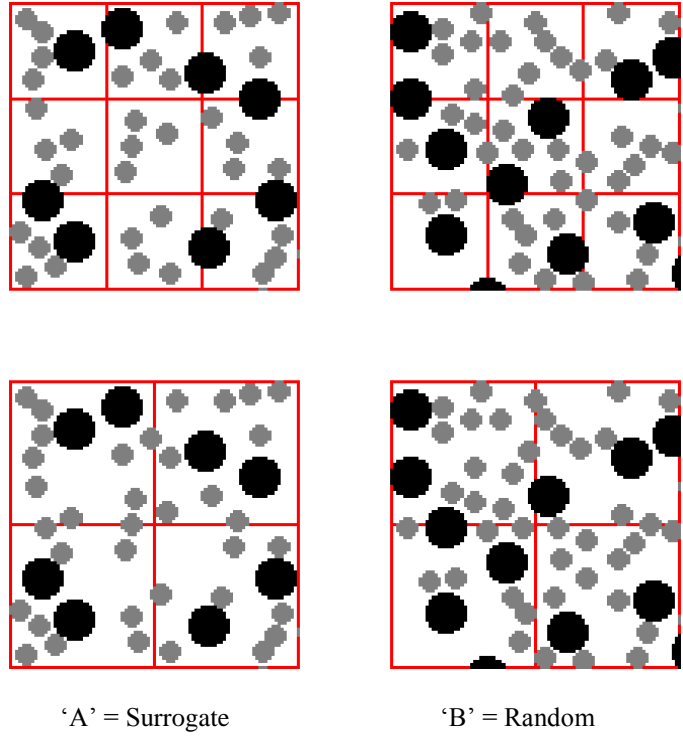

'A' = Surrogate        'B' = Random

**Fig. 3d**

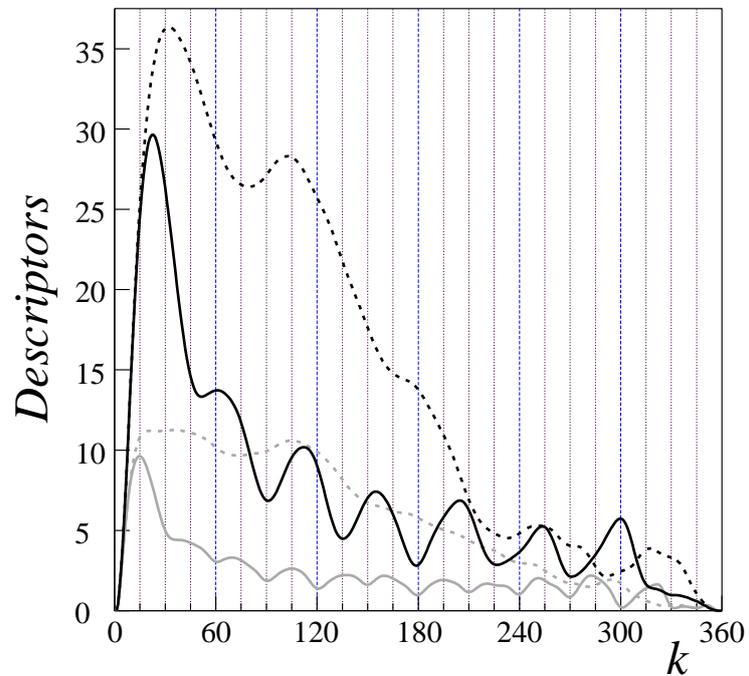

**Fig. 3.** Detection of dissimilar hidden statistical periodicities on example of two patterns, each of size $360 \times 360$, with subtle differences in their arrangement. (a) Surrogate configuration 'A'; (b) Random case 'B'. (c) Only in the case 'A' each cell of size $45 \times 45$ contains exactly two black clusters while four gray clusters occupy each of $30 \times 30$ cell. (d) The contribution of phase descriptors $f_{1,A}$ and $f_{2,A}$ related to black and grey phase, solid lines for the surrogate pattern 'A' while dashed lines correspond to random case 'B'. As expected, only in the case with hidden periodicities, i.e. 'A', appear quite regular intervals between the minima.